\documentclass[dvips,12pt]{article}
\usepackage{epsfig}
\usepackage{graphicx}
\input{psfig}
\def\Journal#1#2#3#4{{#1} {\bf #2}, #3 (#4)}

\def\PRL{\em Phys. Rev. Lett.}

\def\ra{\rightarrow}

\def\be{\begin{equation}}
\def\ee{\end{equation}}
\def\bea{\begin{eqnarray}}
\def\eea{\end{eqnarray}}

\def\etal{{\it et al}}
\begin{document}
\pagenumbering{roman}

\renewcommand{\thefootnote}{\arabic{footnote}}
\pagenumbering{roman}

\begin{center}
{\Large\bf BTeV -  an Experiment to Measure Mixing, CP Violation
and Rare Decays of Beauty and Charm at the Fermilab
Collider\footnote{Submitted to the XXXth International Conference on High 
Energy Physics, July 27 - August 2, 2000, Osaka, Japan}
}
\end{center}

\begin{abstract}
We discuss the physics goals and rationale for a detector to study
Beauty and Charm decays in the forward direction at the Fermilab Tevatron.
We then describe the BTeV detector which has been designed 
to achieve these goals and present its physics reach based on extensive 
simulation. We conclude by comparing BTeV to other experiments designed to 
explore similar topics.
\end{abstract}

\vspace{0.2in}

\begin{center}
{\Large The BTeV Collaboration}
\end{center}
\normalsize

\begin{tabbing}
xxxx \= fill up the rest of the space with stuff \kill
A. Kulyavtsev, M. Procario\footnote{Currently at US Department of Energy},
J. Russ, and J. You \\
 \> Carnegie Mellon University, Pittsburgh, PA 15213, USA \\
J. Cumalat, University of Colorado, CO 80309, USA \\
J. A. Appel, C. N. Brown, J. Butler, H. Cheung, D. Christian, G. Chiodini, \\
S. Cihangir, I. Gaines, P. Garbincius,  L. Garren, E. E. Gottschalk, \\
G. Jackson, P. Kasper, P. H. Kasper, R. Kutschke,  S. W. Kwan, P. Lebrun,\\
P. McBride, L. Stutte, and J. Yarba, Fermilab, Batavia, IL 60510, USA \\
P. Avery, and M. Lohner,  University of Florida, Gainesville, FL 32611, USA \\
R. A. Burnstein, D. M. Kaplan, L. M. Lederman, H. A. Rubin, and C. White \\
 \> Illinois Institute of Technology, Chicago, IL 60616, USA \\
D. Kim, M. Selen, and J. Wiss, University of Illinois at Urbana-Champaign,\\ 
 \> IL 61801-3080, USA \\
R. W. Gardner and D. R. Rust, Indiana University, Bloomington, Indiana, 47405, USA\\
E. Casimiro, D. Menasce, L. Moroni, D. Pedrini, and S. Sala \\
\> INFN and University of Milano, Italy \\
G. Boca, G. Liguori, and P. Torre, Dipartimento di Fisica Nucleare e Teorica,\\
\> Universita' di Pavia and INFN, Sez. di Pavia, Italy \\
A. A. Derevschikov, Y. M. Goncharenko, V. Yu. Khodyrev, A. P. Meschanin, \\ 
L. V. Nogach, K. E. Shestermanov, L. F. Soloviev, and A. N. Vasiliev\\
\>Institute of High Energy Physics (IHEP), Protvino, Moscow Region, Russia \\
C. Newsom, The University of Iowa, Iowa City, IA 52242-1479 \\
Y. Kubota, R. Poling, A. Smith and B. Speakman, University of Minnesota \\ 
\> MN 55455, USA  \\
T. Y. Chen, Nanjing University, Nanjing 210008, China \\
V. Papavassiliou, New Mexico State University  \\
K. Honscheid, and H. Kagan, Ohio State University, Columbus, OH 43210, USA \\
W. Selove, University of Pennsylvania, Philadelphia, PA 19104, USA \\
A. Lopez , University of Puerto Rico, Mayaguez, Puerto Rico \\
T. Coan, Southern Methodist University, Dallas, TX 75275, USA \\
M. Alam, State University of New York at Albany, Albany, NY 12222, USA  \\
X. Q. Yu, University of Science and Technology of China, \\
\> Joint Institute for High Energy Physics, Hefei, Anhui 230027, China \\
M. He, Shandong University, Jinan, Shandong 250100, China  \\
S. Shapiro (emeritus), Stanford Linear Accelerator Center, PO Box 4349, \\
\> Stanford, CA 94309, USA \\
M. Artuso, G. Majumder, R. Mountain, T. Skwarnicki, S. Stone, J. C. Wang, and 
A. Wolf \\
\> Syracuse University, Syracuse, NY 13244-1130, USA \\
K. Cho, T. Handler and R. Mitchell, University of Tennessee \\
\> Knoxville, TN 37996-1200, USA \\
A. Napier, Tufts University, Medford, MA 02155, USA \\
D. D. Koetke, Valparaiso University, Valparaiso, IN 46383, USA  \\
W. Johns, P. Sheldon, K. Stenson, and M. Webster, Vanderbilt University, \\
 \> Nashville, TN 37235, USA \\
M. Sheaff, University of Wisconsin, Madison, WI 53706, USA \\
J. Slaughter, Yale University, High Energy Physics,New Haven, CT 06511, USA \\
S. Menary, York University,  ON M3J 1P3, Canada \\
\end{tabbing}

\newpage

\pagenumbering{arabic}
\setcounter{figure}{0}

\section{Motivation}
BTeV \cite{proposal} is a program designed to challenge the Standard Model 
explanation of
CP Violation, mixing and rare decays in the $b$ and $c$ quark systems.
Exploiting the large number of $b$'s and $c$'s produced at the Tevatron collider,
we will make precise measurements of Standard Model parameters and
an  exhaustive search for physics beyond the Standard Model.

BTeV can perform the compelling physics studies that need to be done, 
and is not limited by current constraints on what studies can be done.
We are not constrained by a central geometry that is prescribed
to study high $p_t$ physics, nor are we limited by relatively low numbers
of $b$-flavored hadrons as in $e^+e^-$ colliders.  BTeV
excels in several crucial areas including: triggering on decays with purely
hadronic final states,  charged particle identification, electromagnetic
calorimetry and proper time resolution.

In the Standard Model, CP violation has its origin in the phenomenon of
quark mixing. As a result, the Standard Model makes very specific connections
among various kinds of CP violating $B$ decays and among $B$ decays,
kaon and charm decays. Standard Model quark mixing is described
by the Cabibbo-Kobayashi-Maskawa matrix \cite{ckm},
\begin{equation}
\left(\begin{array}{c}d'\\s'\\b'\\\end{array} \right) =
\left(\begin{array}{ccc} 
V_{ud} &  V_{us} & V_{ub} \\
V_{cd} &  V_{cs} & V_{cb} \\
V_{td} &  V_{ts} & V_{tb}  \end{array}\right)
\left(\begin{array}{c}d\\s\\b\\\end{array}\right)~~.
\end{equation}
The unprimed states are the mass eigenstates, while the primed states denote
the weak eigenstates. The $V_{ij}$'s are complex numbers that can be
represented by four independent real quantities, if the matrix is unitary.
These numbers are  fundamental constants of nature that 
need to be determined from experiment, as with any other
fundamental constant such as $\alpha$ or $G$.
Measuring them accurately is important, but the most important goal of
BTeV is to make a broad range of measurements to check whether the whole
picture is correct. If inconsistencies appear, that means there is 
new physics in play, physics beyond the Standard Model. More detailed 
study would then elucidate the nature of this new physics.


To confront the Standard Model,
measurements are necessary on CP violation in $B^o$ and $B_s$ mesons, $B_s$
mixing, rare $b$ decay rates, and on mixing, CP violation 
and rare decays in the charm sector.

Although much has been learned about $b$ and $c$ decays
from past and current experiments, and more will be learned soon; many, if
not most, of the  crucial 
measurements will not have been made by the dawn of the LHC era. 
It is just
as important to see if the ``Standard Model'' explains quark mixing and CP 
violation as it is to see if there is a ``Standard Model"
 Higgs particle which generates mass.

With BTeV, we can mount a formidable assault on the CKM explanation of 
CP violation and mixing. Simply stated, we must do this physics!

\section{Physics Goals}


BTeV is designed to make a complete enough set of measurements on the decays 
of hadrons containing $b$ and $c$ quarks so as to be able to either accurately
determine Standard Model parameters or to discover fundamental 
inconsistencies that could lead us to an understanding beyond the model.
The most important measurements to make involve
 mixing, CP violation and
rare decays of hadrons containing $b$ or $c$ quarks.


Using unitarity, Aleksan, Kayser and London \cite{KAL} have shown that the CKM matrix can
be expressed in terms of four independent phases. These
are taken as:
\begin{eqnarray}
\beta=arg\left(-{V_{tb}V^*_{td}\over V_{cb}V^*_{cd}}\right),&~~&
\gamma=arg\left(-{{V^*_{ub}V_{ud}}\over {V^*_{cb}V_{cd}}}\right), \nonumber\\
\chi=arg\left(-{V^*_{cs}V_{cb}\over V^*_{ts}V_{tb}}\right),&~~&
\chi'=arg\left(-{{V^*_{ud}V_{us}}\over {V^*_{cd}V_{cs}}}\right)~~. 
\end{eqnarray}
Another phase $\alpha$, the angle between $V_{ub}$ and $V_{td}$, is 
redundant with $\beta$ and $\gamma$, since
\begin{equation}
\alpha+\beta+\gamma = \pi~~~.
\end{equation}

It is important to uniquely measure all of these phases, including $\alpha$.
CP asymmetry measurements often involve measuring  $\sin(2\phi)$, where
$\phi$ is the angle of interest.
When we measure $\sin(2\phi)$ we have a four-fold 
ambiguity in $\phi$, namely $\phi$, $\pi/2-\phi$, $\phi+\pi$ and $3\pi/2-
\phi.$ These ambiguities can mask the effects of new physics. 
One of our main  tasks is to
remove as many of the ambiguities as possible.

A complete program includes measuring the CP violating angles 
$\alpha,~\beta,~\gamma$ and $\chi$, measuring the $B_s$ oscillation
frequency, searching for anomalous rates in ``rare" $b$ decays and searching
for mixing and CP violation in the charm sector, where Standard Model 
rates are expected to be small and new physics could have large signals.

The ``Physics Case," presented in Chapter 1 of the proposal, 
describes in detail the measurements
we wish to make and the specific decay modes that we envision using.
Table~\ref{table:reqmeas} lists the most important physics quantities and
suggested decay modes which measure them. We also list the detector 
characteristics needed to make each measurement. 
(Rare $b$ decay measurements and charm physics are not included in this table.)

\begin{table}[tbp]
\begin{center}
\caption{Required CKM measurements for $B$ mesons and associated key detector
characteristics.}
\label{table:reqmeas}
\begin{tabular}{llcccc} \hline\hline
Physics & Decay Mode & Hadron & $K\pi$ & $\gamma$ & Decay \\
Quantity&            & Trigger & Sep   & Det & Time $\sigma$ \\
\hline
$\sin(2\alpha)$ & $B^o\to\rho\pi\to\pi^+\pi^-\pi^o$ & $\surd$ & $\surd$& $\surd$ 
&\\
$\cos(2\alpha)$ & $B^o\to\rho\pi\to\pi^+\pi^-\pi^o$ & $\surd$ & $\surd$& 
$\surd$ &\\
sign$(\sin(2\alpha))$ & $B^o\to\rho\pi$ \& $B^o\to\pi^+\pi^-$ & 
$\surd$ & $\surd$ & $\surd$ & \\
$\sin(\gamma)$ & $B_s\to D_s^{\pm}K^{\mp}$ & $\surd$ & $\surd$ & & $\surd$\\
$\sin(\gamma)$ & $B^-\to \overline{D}^{0}K^{-}$ & $\surd$ & $\surd$ & & \\
$\sin(\gamma)$ & $B^o\to\pi^+\pi^-$ \& $B_s\to K^+K^-$ & $\surd$ & $\surd$& & 
$\surd$ \\
$\sin(2\chi)$ & $B_s\to J/\psi\eta',$ $J/\psi\eta$ & & &$\surd$ &$\surd$\\
$\sin(2\beta)$ & $B^o\to J/\psi K_S$ & & & & \\
$\cos(2\beta)$ &  $B^o\to J/\psi K^o$, $K^o\to \pi\ell\nu$  & & & & \\
$\cos(2\beta)$ &  $B^o\to J/\psi K^{*o}$ \& $B_s\to J/\psi\phi$  & & & 
&$\surd$ \\
$x_s$  & $B_s\to D_s^+\pi^-$ & $\surd$ & & &$\surd$\\
$\Delta\Gamma$ for $B_s$ & $B_s\to  J/\psi\eta'$, $ D_s^+\pi^-$, $K^+K^-$ &
$\surd$ & $\surd$ & $\surd$ & $\surd$ \\
\hline
\end{tabular}
\end{center}
\end{table}

The BTeV detector, described below and in Part 2 of the proposal, 
possesses all of the properties
required to carry out these measurements. 
Perhaps just as importantly, the detector is powerful enough
to pursue physics in many areas of $b$ and $c$ production and decay.
In the future, new final states that will be important to measure will
surely emerge. 
BTeV, because of its excellent trigger, tracking, particle
identification and photon detection, will be in prime position to
investigate any such new ideas. 

\section{Rationale for a Forward Detector at the Tevatron}

BTeV covers the forward direction, 10-300 mrad, with respect to both colliding
beams. In Chapter 2 of the proposal we explain the reasons for 
this choice. We summarize them here.

Measured $b\overline{b}$ cross-sections at the Tevatron integrate to
100 $\mu$b \cite{bcx}. One measurement by D0 in the forward region
normalizes to 180 $\mu$b \cite{bcxD0}. 
Conservatively, we use the 100 $\mu$b value
for our physics projections. The yield of $b$-flavored hadrons then is
$\approx 4\times 10^{11}$ in $10^7$ seconds at a luminosity of
$2\times 10^{32}$ cm$^{-2}$s$^{-1}$ and much of it is in the
forward direction. The charm yield is approximately
one order of magnitude higher and even more of it is concentrated in the 
forward direction.

\begin{figure}[tbp]
\vspace{-2.8cm}
\centerline{\psfig{figure=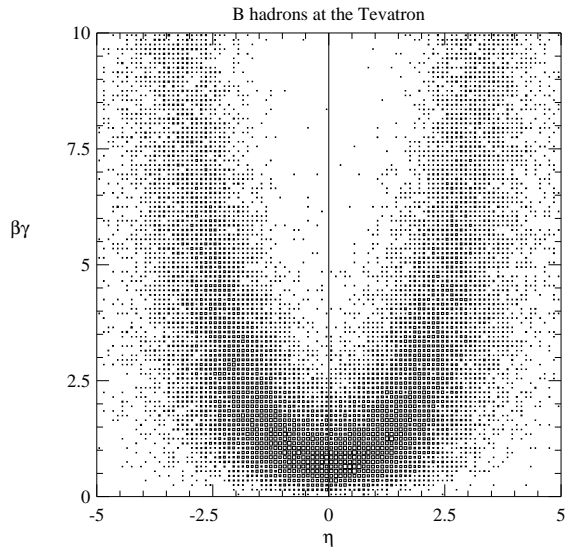,height=4in,bbllx=0bp,bblly=0bp,bburx=600bp,bbury=700bp,clip=}}
\vspace{-.2cm}
\caption{\label{bg_vs_etaex}$\beta\gamma$ of the  $B$  versus $\eta$.}
\end{figure}

According to QCD  calculations of $b$ quark production, 
there is a strong correlation between the $B$ momentum and pseudorapidity,
$\eta$. Shown in
Fig.~\ref{bg_vs_etaex} is the $\beta\gamma$ of the $B$ hadron versus $\eta$,
as computed by the Monte Carlo physics generator 
Pythia at $\sqrt{s}\,=\,2$ TeV.
It can clearly be seen that near $\eta$ of zero, $\beta\gamma\approx 1$, while
at larger values of $|\eta |$, $\beta\gamma$ can easily reach values of ~6.
This is important because the mean decay length varies with $\beta\gamma$
and, furthermore, the absolute momenta of the decay products are larger, 
allowing for a suppression of the multiple scattering error.

\begin{figure}[htb]
\vspace{-.2cm}
\centerline{\psfig{figure=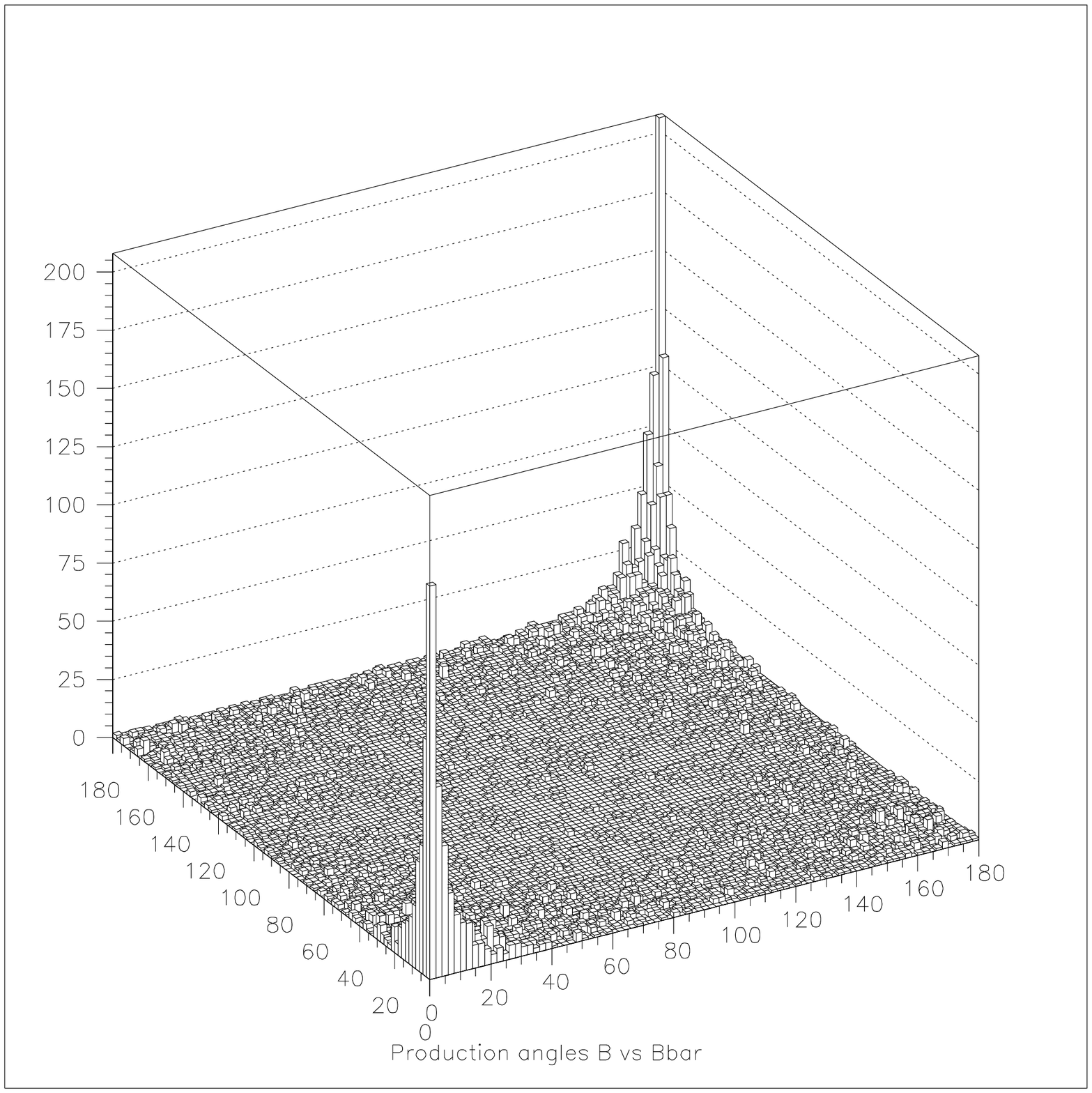,height=4.2in,bbllx=0bp,bblly=0bp,bburx=600bp,bbury=700bp,clip=}}
\vspace{-1.2cm}
\caption{\label{bbar}The production angle (in degrees) for the hadron
containing a $b$ quark plotted versus the production angle for a hadron
containing a $\bar{b}$ quark, from the Pythia Monte Carlo generator.}
\end{figure}

A crucially important correlation of
$b\bar{b}$ production at hadron colliders is shown in Fig.~\ref{bbar},
where  the production
angle of the hadron containing the $b$ quark is plotted versus the production
angle of the hadron containing the $\bar{b}$ quark. Here zero degrees
represents the direction of the incident proton and 180 degrees, the
incident antiproton.  There is a very strong
correlation in the proton or the antiproton directions: 
when the $B$ is forward,
the $\overline{B}$ is also forward. (We call both the proton and antiproton
directions forward.) This correlation between $B$ and $\overline{B}$
production is not present in the central region (near 90 degrees). 

Thus, when a $b$-flavored hadron is produced forward, the accompanying
$\bar{b}$ is also produced in the forward direction, allowing for reasonable
levels of flavor tagging. The large $b$ quark yield, the long $B$ decay
length,  the correlated
acceptance for both $b$'s and the suppression of multiple scattering
errors due to the high $b$ momenta, make the forward direction
an ideal choice.

\section{Detector Description}

\begin{figure}[tbp]
\centerline{\epsfig{figure=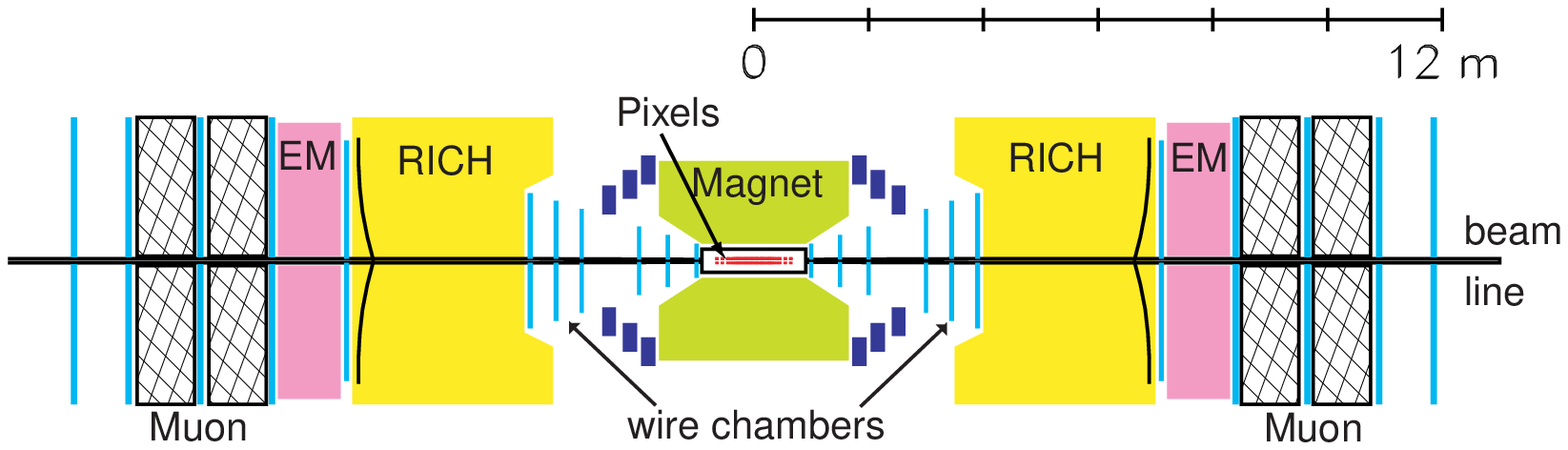,height=1.9in}}
\caption{\label{btev_det_simp}A sketch of the BTeV detector. The two 
arms are identical.}
\end{figure}

A sketch of the detector is shown in Fig.~\ref{btev_det_simp}.
The geometry is complementary
to  that used in current collider experiments. 
The detector looks similar to a fixed target experiment, but
has two arms, one along the proton direction and the other along the
antiproton direction.

The key design
features of BTeV include:
\begin{itemize}
\item A dipole located on the IR, which  gives BTeV an effective ``two arm''
           acceptance;
\item A precision vertex detector based on planar pixel arrays;
\item A detached vertex trigger at Level 1 that makes BTeV efficient for
   most final states, including purely hadronic modes;
\item Excellent particle identification using a Ring Imaging Cherenkov Detector
(RICH);
\item A high quality PbWO$_4$ electromagnetic calorimeter capable of
reconstructing
final states with single photons, $\pi^o$'s, $\eta$'s or $\eta'$'s,
and of identifying electrons;
\item Precision tracking using straw tubes and silicon microstrip detectors,
which provide excellent momentum and mass resolution;
\item Excellent identification of muons using a dedicated detector with the
ability to supply a dimuon trigger; and
\item A very high speed and high throughput data acquisition system which
eliminates the need to tune the experiment to specific final states.
\end{itemize}                  

Each of these key elements of the detector is discussed in Part 2 of the 
proposal. Here we discuss them briefly.

\subsection{Dipole Centered on the Interaction Region}

A large dipole magnet, with a 1.6 T central field, is centered on the
interaction region.
In addition to giving us a
compact way of providing momentum measurements in  both ``forward'' 
directions, it provides magnetic
deflection in the vertex detector, which is exploited by the trigger to
remove low momentum tracks, which could have been deflected by multiple Coulomb
scattering, from its search for detached tracks.

\subsection{The Pixel Vertex Detector}

In the center of the magnet there is a silicon pixel vertex detector.
This detector serves two functions: 
it is 
an integral part of the charged particle tracking system, providing
accurate vertex information for the offline analysis; and 
it delivers very clean, precision space points to the BTeV vertex trigger.

\begin{figure}[tbp]
\centerline{\epsfig{file=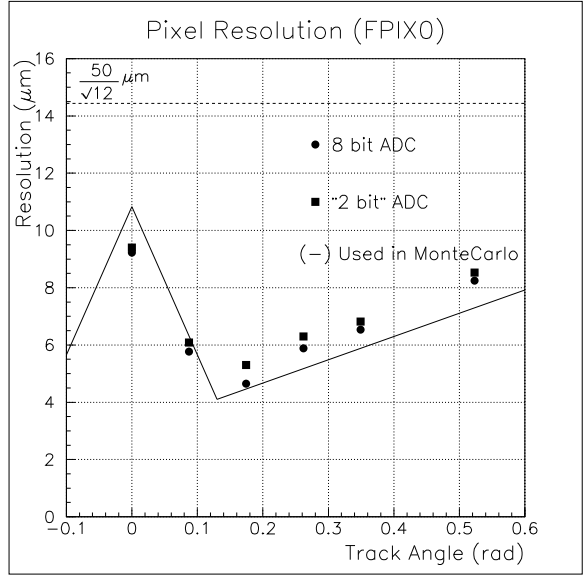, height=3.0in}}
\caption{\label{pixel_res}The resolution achieved in our test beam run
using 50 $\mu$m wide pixels and an 8-bit ADC (circles) or a 2-bit ADC 
(squares), compared with our simulation (line).}
\end{figure}

We have tested prototype pixel devices in a beam at Fermilab. These consist
of \hbox{50 $\mu$m $\times$ 400 $\mu$m} pixels bump-bonded to custom made 
electronics chips developed at Fermilab. A comparison of the
position resolution achieved in the test beam and the Monte Carlo simulation
is shown in Fig.~\ref{pixel_res}.
The resolution is excellent and exceeds our  requirement of 9~$\mu$m.

The critical quantity for a $b$ experiment is $L/\sigma_L$, where $L$ is
the distance between the primary (interaction) vertex and the secondary (decay)
vertex, and $\sigma_{L}$ is its error.
For central detectors
the $B$'s are slower, because the mean transverse $B$ momentum is 5.3 GeV/c,
virtually independent of the longitudinal momentum. Since they also suffer more
multiple scattering, they have relatively poorer $L/\sigma_L$ distributions.
LHC-b, on the other hand, does not benefit by going to higher momentum because,
after a momentum of around 10 GeV/c (depending on the detector),
$\sigma_L$ also increases linearly.

\begin{figure}[tbp]
\vspace{-1.5cm}
\centerline{\epsfig{file=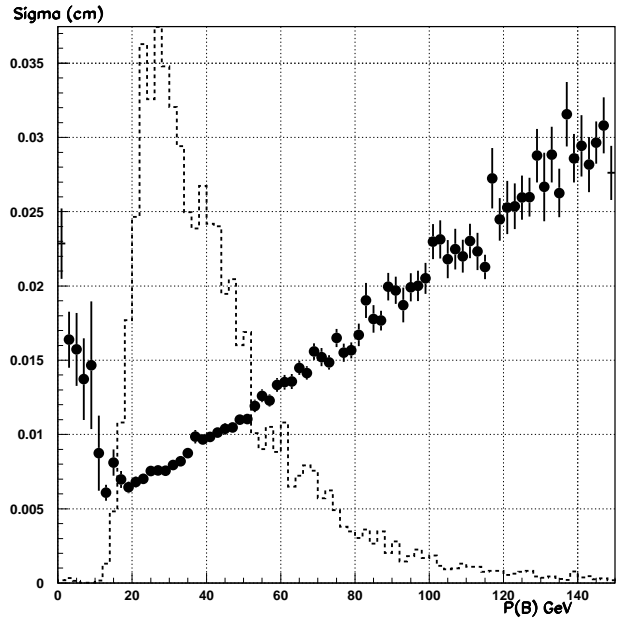, height=4.0in}}
\caption{\label{das_pipi}The $B$ momentum distribution for
$B^o\to\pi^+\pi^-$ events (dashed) and the error in decay length
$\sigma_L$ as a function of momentum.}
\end{figure}

The efficacy of this geometry is illustrated by considering the distribution
of the resolution on the $B$ decay length, $L$, for the decay 
$B^o\to\pi^+\pi^-$.
Fig.~\ref{das_pipi} shows
the r.m.s. errors in the decay length as a function of
momentum;  it also shows the momentum distribution of the $B$'s accepted
by BTeV.
The following features are noteworthy:
\begin{itemize}
\item The $B$'s used by BTeV peak at $p$ = 30 GeV/c and average about 40 GeV/c.
\item The decay length is equal to $450~\mu$m $\times\, p/M_B$.
\item The error on the decay length is smallest near the peak of our
accepted momentum distribution. It increases at lower values of $p$, due to
multiple scattering, and increases at larger values of $p$ due to the 
smaller angles of the Lorentz-boosted decay products. 
\end{itemize}

\subsection{The Detached Vertex Trigger}

It is impossible to record data from each of the 7.5 million
beam crossings per second. A prompt decision, colloquially called a 
``trigger,'' must be made to record or discard the data from each crossing.
The main BTeV trigger is provided by the silicon pixel detector.
The Level 1 Vertex Trigger inspects every beam crossing and, using only
data from the pixel detector, reconstructs the primary vertices and
determines whether there are detached tracks which could signify a $B$
decay. Since  the $b$'s are at high momentum, 
the multiple scattering of the decay products is minimized allowing for
triggering on detached heavy quark decay vertices.

With our outstanding pixel resolution, we are able to trigger efficiently 
at Level 1 on a variety
of $b$ decays. The trigger has been fully simulated, including the pattern
recognition code. In Table~\ref{tab:trigeff} we give the efficiencies to trigger on a sample
of final states providing that the particles are in the detector acceptance
and otherwise pass all the analysis cuts. We see the trigger efficiencies
are generally above 50\% for the $b$ decay states of interest and at
the 1\% level for minimum bias background. These numbers are
evaluated at an average rate of 2 interactions per beam crossing,
corresponding to our design luminosity of 2$\times 10^{32}$ cm$^{-2}$s$^{-1}$.

\begin{table}
\caption{Level 1 trigger efficiencies for minimum-bias events and various
         processes of interest that are required to pass offline analysis cuts.
         All trigger efficiencies are determined
         for beam crossings with an average of two interactions per crossing
         using the Monte Carlo code shown in the table.}
\label{tab:trigeff}
\vspace{3mm}
\centering

\begin{tabular}{lcc}
\hline\hline
Process & Eff. (\%) & Monte Carlo \\
\hline\hline
Minimum bias & 1 & BTeVGeant\\
\hline
$B_s\to D_s^+ K^-$        & 74 & BTeVGeant \\
$B^0\to D^{*+} \rho^-$     & 64 & BTeVGeant\\
$B^0\to \rho^0 \pi^0$ & 56 & BTeVGeant \\
$B^0\to J/\psi K_s$   & 50 & BTeVGeant \\
$B_s\to J/\psi K^{*o}$   & 68 & MCFast\\
$B^-\to D^0 K^- $     & 70 & MCFast\\
$B^-\to K_s \pi^- $   & 27 & MCFast\\
$B^0\to$ 2-body modes & 63 & MCFast\\
$(\pi^+\pi^-, K^+\pi^-,K^+K^-)$ & & \\
\hline\hline
\end{tabular}

\end{table}

\subsection{Charged Particle Identification}

Charged particle identification is an absolute requirement for a modern
experiment designed to study the decays of $b$ and $c$ quarks. 
The relatively open
forward geometry has sufficient space to install a Ring Imaging
Cherenkov detector (RICH), which provides powerful particle ID 
capabilities over a broad range of momentum.
The BTeV RICH detector must separate pions from kaons and protons in a momentum
range from $3-70$~GeV/$c$. The lower momentum limit
is determined by soft kaons useful
for flavor tagging, while the higher momentum limit
is given by two-body $B$ decays.
Separation is accomplished using a gaseous freon radiator
to generate Cherenkov light in the optical frequency range. The light is then
focused from mirrors onto Hybrid Photo-Diode (HPD) tubes. 
To separate kaons from
protons below 10 GeV/c an aerogel radiator will be used. 

\begin{figure}[tbp]
\vspace{-0.8cm}
\centerline{\epsfig{file=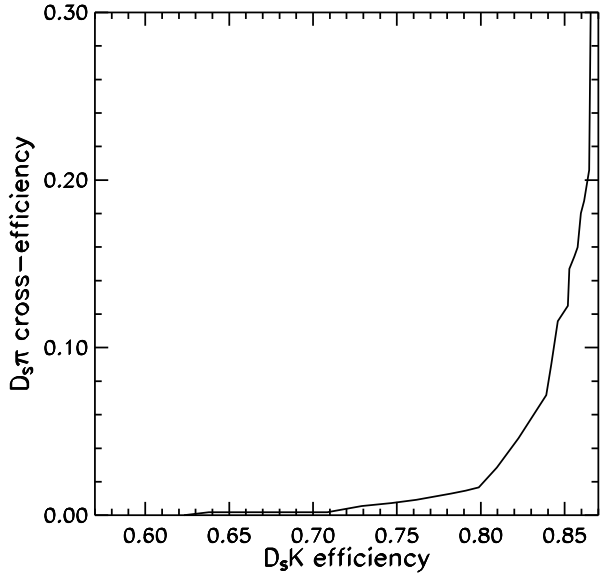, height=3.6in}}
\caption{The efficiency to detect the fast $K^-$ in the
reaction $B_s\to D_s^+K^-$ versus the rate to misidentify the $\pi^-$ from
$B_s\to D_s^+\pi^-$ as a $K^-$.\label{dskvsdspi}}
\end{figure}

As an example of the usefulness of this device we show, in
Fig.~\ref{dskvsdspi},  the efficiency for
detecting the $K^-$ in the decay $B_s\to D_s^+K^-$ versus the rejection for
the $\pi^-$ in the decay $B_s\to D_s^+\pi^-$.  We see that high 
efficiencies can be obtained with excellent rejections. 

\subsection{Electromagnetic Calorimeter}

In BTeV, photons and electrons are detected when they create an electromagnetic
shower cascade in crystals of PbWO$_4$, a dense and transparent medium that 
produces scintillation light.
The amount of light is proportional to the incident
energy. The light is sensed by photomultiplier tubes (or possibly hybrid
photodiodes). The crystals are 22 cm long and have a small transverse 
cross-section, 
26 mm$\times$ 26 mm, providing excellent segmentation.
The energy and position resolutions are exquisite,
\begin{eqnarray}
\frac{\sigma_{E}}{E} &= & \sqrt{{{(1.6\%)^2}\over{E}}+(0.55\%)^2}~, \\
\sigma_x &=& \sqrt{{{(3500~\mu m)^2}\over{E}}+(200~\mu m)^2}~,
\end{eqnarray}
where $E$ is in units of GeV. This leads to an r.m.s. $\pi^o$ mass
resolution between 2 and 5 MeV/$c^2$ over the $\pi^o$ momentum range 1 to 40 GeV/c.

The crystals are designed to point at the center of the interaction region.
They start at a radial distance of 10 cm with respect to the beam-line and
extend out to 160 cm. They cover $\sim$210 mrad. This is smaller
than the 300 mrad acceptance of the tracking detector; 
the choice was made to reduce costs. For
most final states of interest this leads to a loss of approximately 20\%
in signal. 

\begin{figure}[tbp]
\centerline{\epsfig{file=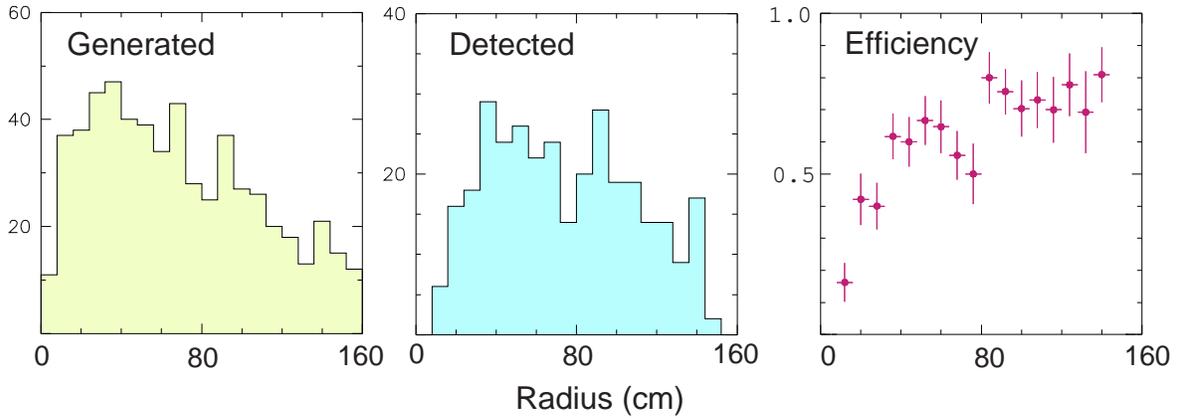, height=2.2in}}
\caption{The radial distribution of generated and detected
photons from $B^o\to K^*\gamma$ and the resulting efficiency. The detector
was simulated by GEANT and the resulting crystal energies were clustered by
our software. The charged tracks from the $K^*$ were required to 
hit the RICH. The simulation was run at 2 interactions/crossing.}
\label{eff_btev_3sigma}
\end{figure}

The calorimeter, at 2 interactions per crossing, has a high rate close to the
beam pipe, where the reconstruction efficiency and resolution
is degraded by overlaps with
other tracks and photons. As we go
out to larger radius, the acceptance becomes quite good. This can be seen
by examining the efficiency of the $\gamma$ in the decay $B^o\to K^*\gamma$,
$K^*\to K^- \pi^+$. Here the decay products of the $K^*$ are required 
to reach the RICH detector. Fig.~\ref{eff_btev_3sigma} shows the radial 
distribution of the generated
$\gamma$'s, the reconstructed $\gamma$'s and the $\gamma$ efficiency versus
radius. The shower reconstruction code, described in Chapter 
12 of the proposal, was developed from that used for the CLEO
CsI calorimeter; for reference, the efficiency of the CLEO barrel 
electromagnetic calorimeter is 89\%.

\subsection{Forward Tracking System}

The other components of the charged-particle tracking system are straw-tube
wire proportional chambers and, near the beam where
occupancies are high, silicon microstrip detectors. These devices are used 
primarily for track momentum
measurement, $K_s$ detection and the Level 2 trigger. These detectors
measure the deflection of charged particles by the BTeV analyzing magnet and
give BTeV excellent mass and momentum resolution for charged particle
decay modes.

\subsection{Muon Detection}

Muon detection is accomplished by insisting that the candidate charged track
penetrate several interaction lengths of magnetized iron and insuring that the
momentum determined from the bend in the toroid matches that 
given by the main spectrometer tracking system. The
muon system is also used to trigger on the dimuon decays of the $J/\psi$. This
is important not only to gather more signal but as a cross check on the
efficiency of our main trigger, the  Detached Vertex Trigger. 

\subsection{Data Acquisition System}

BTeV has a data acquisition system (DAQ) which is capable of recording
a very large number of events. The full rate of $B$'s whose decay products
are in the detector is very high, over 1 kHz. The (direct) charm rate is 
similar. Other experiments are forced by the limitations of their 
data acquisition system to make very harsh decisions on which $B$ events to 
take. BTeV can record nearly all the potentially interesting $B$ and charm 
candidates in its acceptance. Therefore it can address many topics that might
be discarded by an experiment whose DAQ is more restrictive.
Since nature has a way of surprising us, we view the open nature of the
BTeV trigger and the capability of the DAQ as a genuine strength that
offers us the opportunity to learn something new and unanticipated.

\section{Simulation Results and Physics Reach}

The physics reach of BTeV has been established by an extensive and 
sophisticated program of simulations, which is described in detail
in Part 3 of the proposal.
We have simulated the efficiencies and backgrounds in the decay modes used to
measure the CP violating angles $\alpha$, $\beta$, $\gamma$ and $\chi$, the
$B_s$ mixing parameter $x_s$ and a few rare decay final states. 

We have used two simulation packages, GEANT and MCFast. These tools and 
their use are
explained in the beginning of Part III. Briefly, GEANT models all physical
interactions of particles with material and allows us to see the effects 
of hard to calculate backgrounds.  The goal of MCFast is to 
provide a fast, parametrized simulation which is more flexible than GEANT
but not quite as complete in its modeling of physics processes.

In Table~\ref{tab:physics_sum} we give the decay mode, the number of
signal events found in $10^7$ seconds at a luminosity of 
$2\times 10^{32}$ cm$^{-2}$s$^{-1}$ and the signal/background ratio for many
of the interesting decay modes which are possible for BTeV.
We also estimate the error in the relevant physics parameter, if possible,
or the reach in $B_s$ mixing. In some cases more than one reaction
is used to determine a value; in that case they are put between
horizontal lines.

\begin{table}[tbp]
\caption{Summary of physics reach in $10^7$s. 
Pairs of reactions between two lines are
used together.}
\label{tab:physics_sum}
\centering
\begin{tabular}{lrrccc}
\hline\hline
Process &\# of Events & $ S/B$ & Parameter &Error or (Value) \\
\hline\hline
$B^o\to\pi^+\pi^-$ & 24,000 & 3  & Asym. & 0.024  \\
$B_s\to D_s^{\pm} K^{\mp}$ & 13,100 & 7 & $\gamma$ & $7^{\circ}$ \\
$B^o\to J/\psi K_S$ & 80,500 & 10 &$\sin(2\beta)$& 0.025 \\
$B_s\to D_s^+\pi^-$& 103,000&3&$x_s$& (75)\\\hline
$B^-\to \overline{D}^0(K^+\pi^-) K^-$ & 300 & 1 &$\gamma$ & $10^{\circ}$\\
$B^-\to D^0 (K^+K^-) K^-$ & 1,800 & $>$10 &$\gamma$ & 
$10^{\circ}$\\\hline
$B^-\to K_S\pi^-$&8,000 & 1& $\gamma$ & $<5^{\circ}$\\
$B^o\to K^+\pi^-$ &108,000& 20& $\gamma$& $<5^{\circ}$\\\hline
$B^o\to\rho^{\pm}\pi^{\mp}$ & 9,400& 4.1  & $\alpha$ & $\sim 10^{\circ}$ \\
$B^o\to\rho^o\pi^o$ & 1,350& 0.3   & $\alpha$ & $\sim 10^{\circ}$\\\hline
$B_s\to J/\psi\eta$& 1,920& 15& $\sin(2\chi)$& 0.033\\
$B_s\to J/\psi\eta'$&7,280& 30& $\sin(2\chi)$ & 0.033\\\hline
$B^-\to K^-\mu^+\mu^-$&1280& 3.2& &\\
$B^o\to K^*\mu^+\mu^-$&2200&10&&\\
\hline\hline
\end{tabular} 
\end{table}

\begin{figure}
\centerline{\includegraphics[width=6in]{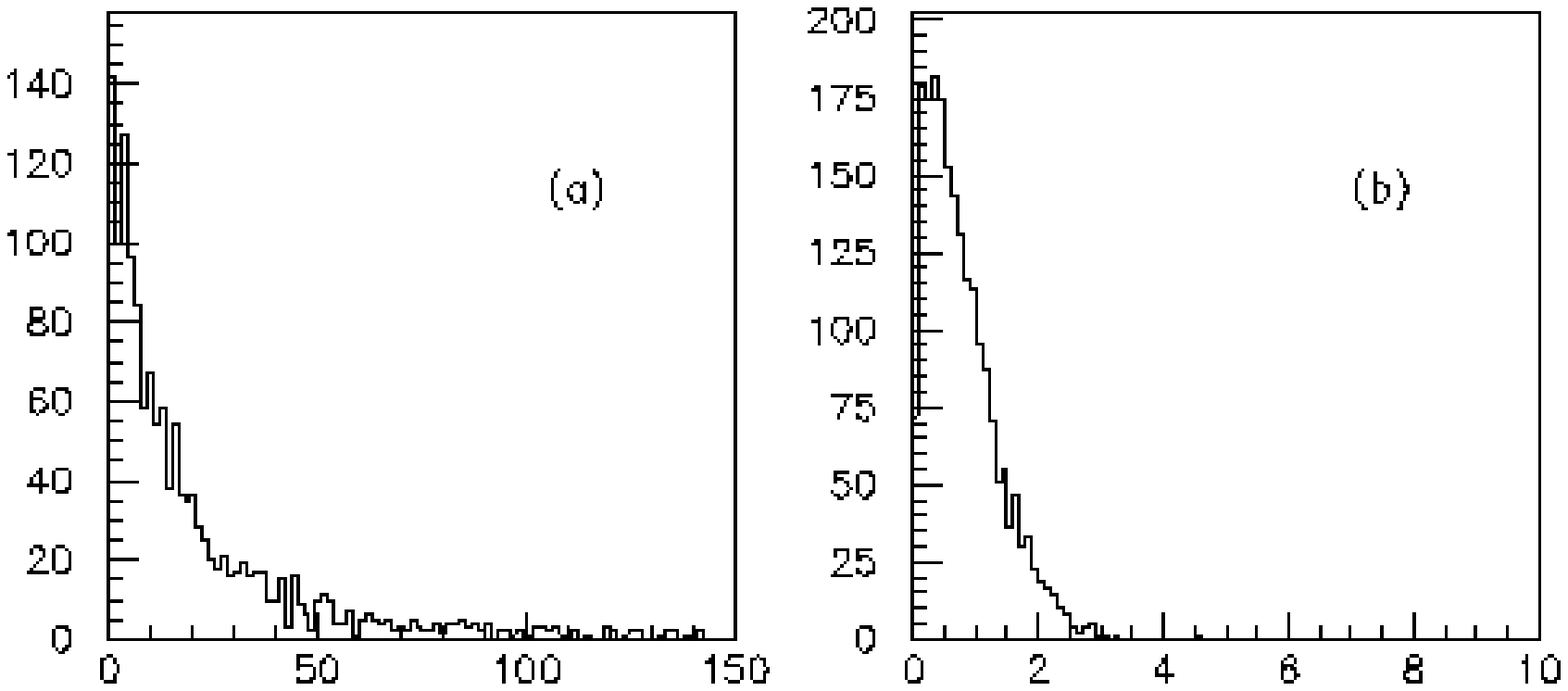}}
\caption{Distributions of $L/\sigma_L$ for (a) $J/\psi$ candidates
         from $B^{o}\rightarrow J/\psi K_{s}$ 
         and (b) prompt $J/\psi$ candidates.  The prompt candidates
         are suppressed by requiring $L/\sigma_L>4$.}
\label{fig:prop00psiks_los}
\centerline{\includegraphics[width=6in]{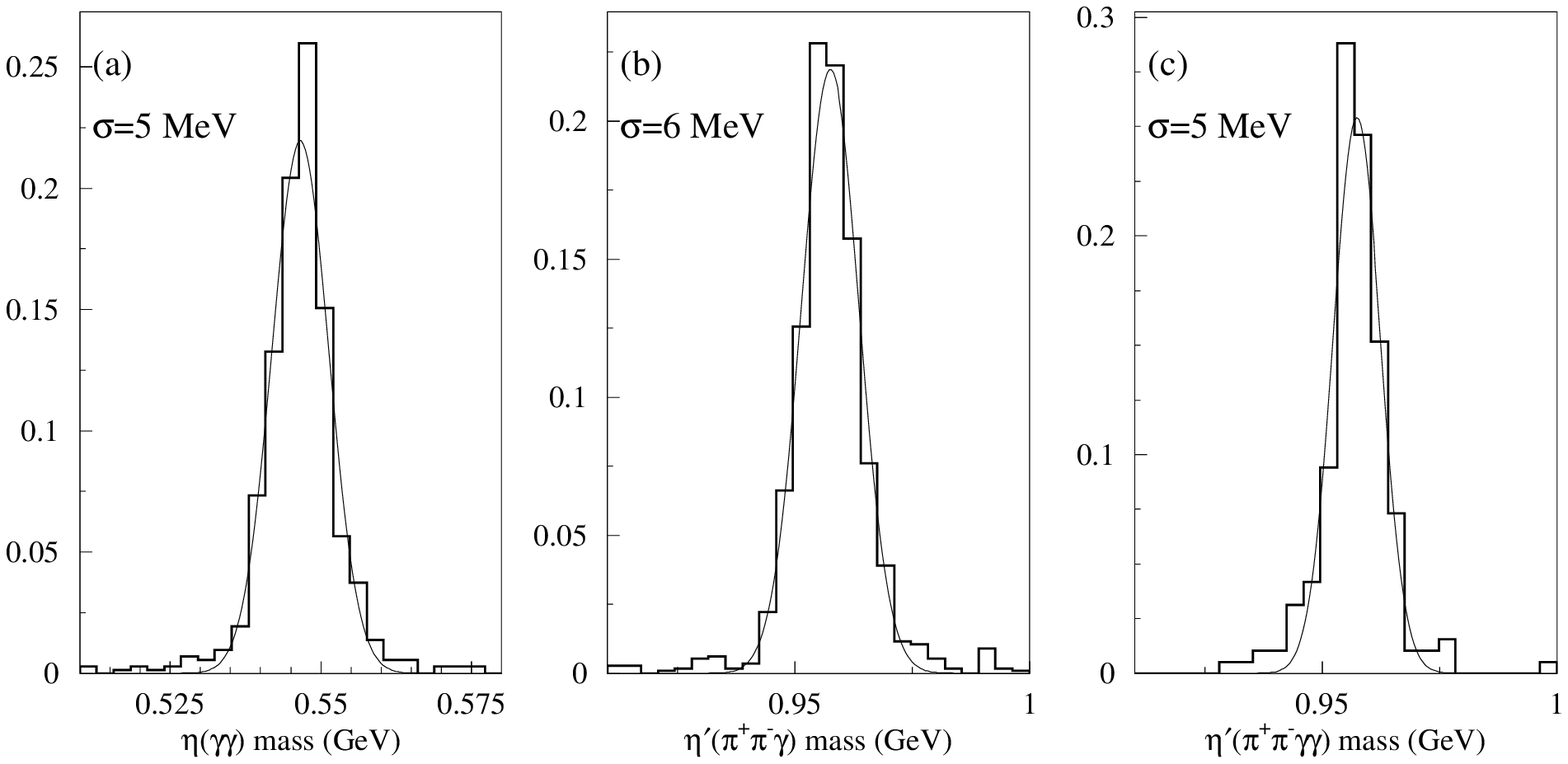}}
\caption{The invariant mass distributions for (a) $\eta\to\gamma\gamma$,
(b) $\eta'\to\pi^+\pi^-\gamma$, and $\eta'\to\pi^+\pi^-\eta$,
$\eta\to\gamma\gamma$. The Gaussian mass resolutions are indicated. 
$\eta$ and $\eta'$ candidates can be used to measure $\chi$ using the
decay mode $B_s \ra J/\psi \eta'$ and $B_s \ra J/\psi \eta$.}
\label{fig:prop00etamass}
\end{figure}

\begin{figure}
\centerline{\includegraphics[height=3.2in,width=6in]{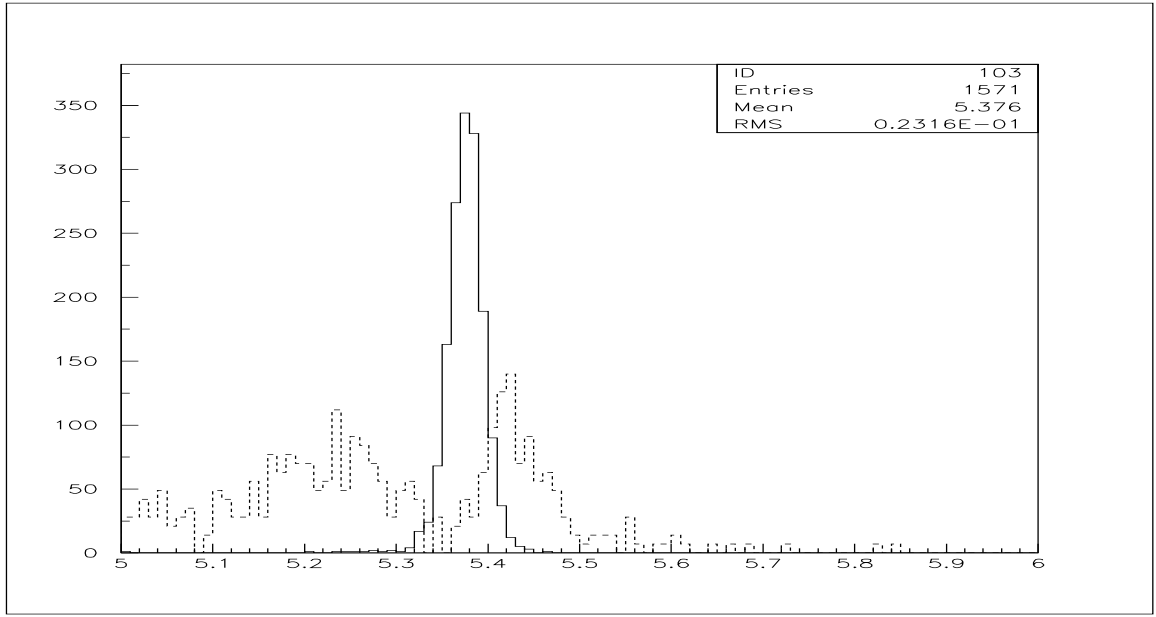}}
\caption{Comparison of $B_s\ra D_s^{+} K^{-}$ signal and background
 from $B_s\ra D_s X$, where X contains at
 least one pion misidentified as a $K^{-}$.  This state is used to 
measure $\gamma$.}
\label{fig:prop00misid}
\end{figure}

\begin{figure}
\centerline{\includegraphics[height=3.2in,width=6in]{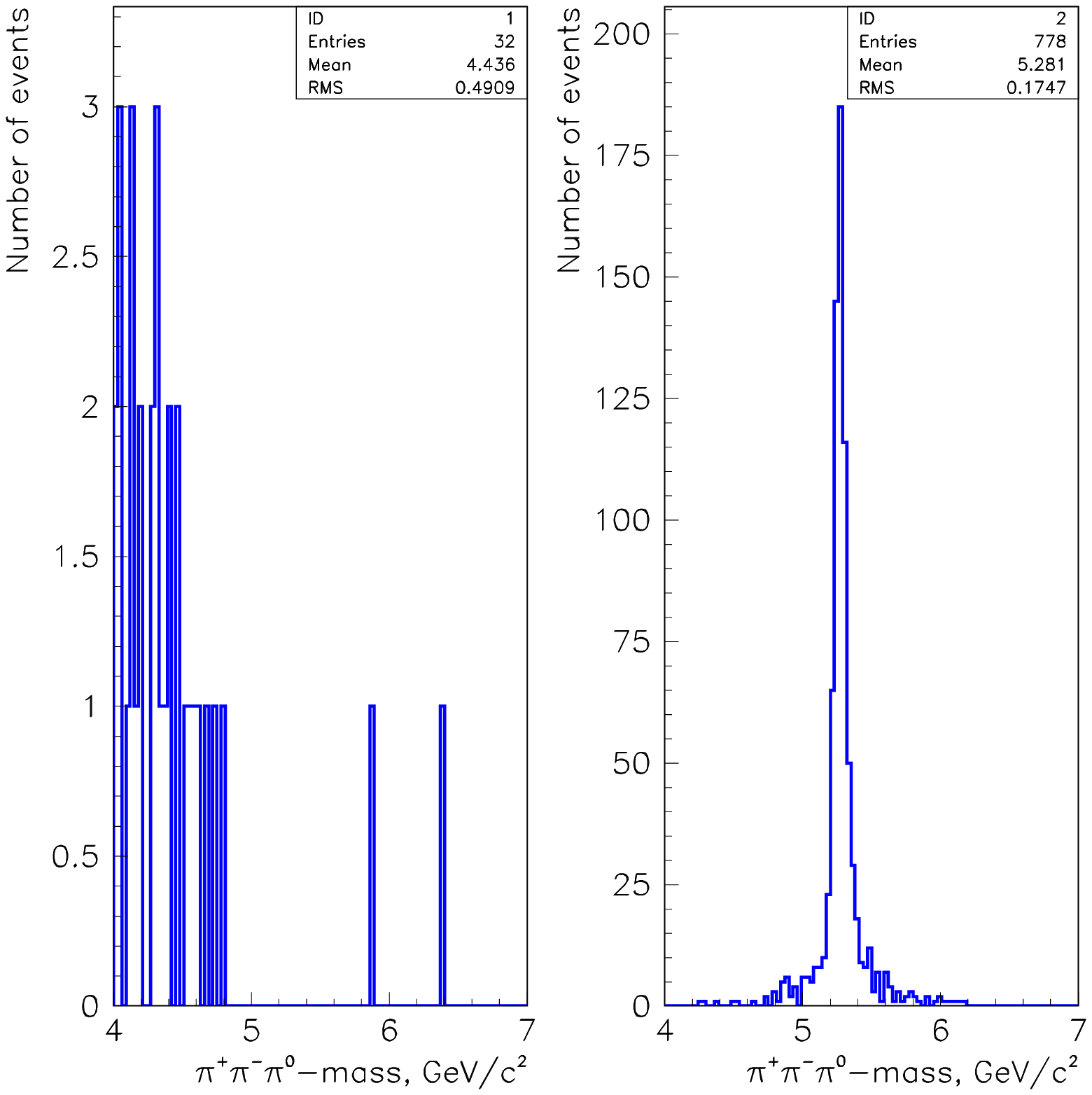}}
\caption{Invariant $\pi^+\pi^-\pi^o$ mass distributions for background
          (left) and signal (right) events for
          $B \rightarrow \rho^+ \pi^-$.  This state is used to 
measure $\alpha$. (The background is not normalized to the signal). }
\label{fig:prop00rhopi}
\end{figure}

Fig.~\ref{fig:prop00psiks_los} shows the power of an $L/\sigma_L$ cut 
on reducing prompt backgrounds in order to extract a clean $B \to J/\psi K_s^o$
signal.  This rejection power comes from the forward geometry combined with
excellent vertex resolution.
Fig.~\ref{fig:prop00etamass} shows the superb capabilities of the 
electromagnetic calorimeter in detecting photons that are used
to reconstruct $\eta$ and $\eta'$ candidates, which can be combined with
$J/\psi$'s to measure $\chi$ through the state $B_s \ra J/\psi \eta^{(')}$.
Fig.~\ref{fig:prop00misid} shows the decay mode $B_s \ra D_s K$ signal
which can be used to measure $\gamma$.  The $\pi^+\pi^-\pi^o$ invariant
mass distributions for $B\rightarrow \rho\pi$ signal and background are 
shown in Fig.~\ref{fig:prop00rhopi}.  The
$B\rightarrow \rho\pi$ decay can be used to measure $\alpha$.

The physics reach of BTeV is extraordinary, even in just one year of running. 
BTeV will be the first to make a precision measurement of the angle $\gamma$;
this will be accomplished in one year of running. The angles $\alpha$ and
$\chi$ will also be measured, though this will take a bit longer. The $B_s$
mixing reach is up to $x_s$ of 75, well above the Standard Model allowed
range of about 40. Thus, not only can BTeV measure the value in the Standard
Model, but it also has a good chance to measure $B_s$ mixing if it is 
determined by physics beyond the Standard Model. Non-Standard Model physics
can also be seen via rare decays where large numbers of reconstructed 
events are expected. 


\section{Comparisons With Other Experiments}
BTeV compares favorably to all other heavy quark experiments.

 BTeV has considerably more reach in every physics
channel than $e^+e^-$ $B$ factories. We give one example here.
Table~\ref{tab:pipiee0} shows a comparison between BTeV and an asymmetric
$e^+e^-$ machine for measuring the CP violating asymmetry in the decay mode
$B^o\to\pi^+\pi^-$. 
It is clear that the large hadronic $b$ production
cross section can overwhelm the much smaller $e^+e^-$ rate. Furthermore, the
$e^+e^-$ $B$ factories do not have access to the important CP violation measurements that
need to be made in $B_s$ decays. Nor can they explore potentially interesting
topics in  the physics of $b$-baryons and $B_{c}$ mesons.

\begin{table}[hbt]
\caption{Number of tagged $B^o\to\pi^+\pi^-$ ($\cal{B}$=$0.43\times 10^{-5}$).}
\begin{center}
\label{tab:pipiee0}
\begin{tabular}{lrccrcr}\hline\hline
&&&&Signal~~~~&Tagging&\\
&$\cal{L}$(cm$^{-2}$s$^{-1})$&$\sigma$&\# $B^o /10^7\,$s&Efficiency&
$\epsilon D^2$& \# Tagged/10$^7\,$s \\
\hline
$e^+e^-$&$3\times 10^{33}$&1.2 nb &$3.6\times 10^{7}$&0.3 &0.3 &13\\
BTeV &$2\times 10^{32}$&100$\mu$b &$1.5\times 10^{11}$&0.037 &0.1
&2370\\\hline\hline
\end{tabular}
\end{center}
\end{table}

CDF and D0 have done useful $b$ physics by triggering on $J/\psi$ decays into 
dimuons. CDF plans more aggressive triggers on selected purely hadronic
final states in the future. However, the kinematics of $b$ decays do not favor the central region. Most of the $b$'s are relatively slow with the peak of the 
transverse momentum distribution being at 5.3 GeV/c, which at an $\eta$ of 1, 
produces $B$'s with $\gamma\beta$ of 1. These relatively slow $B$'s are 
intrinsically difficult to vertex and trigger on. Furthermore, CDF and D0 do not 
have state of the art charged particle identification nor do they possess 
excellent photon detection. 

Although the ATLAS and CMS detectors will have some $b$ physics capabilities,
they will be limited to final states with dileptons or even perhaps only
to $J/\psi$ decays. LHC-b, on the other hand, is an experiment that has been 
designed exclusively to study $b$ decays and provides real competition in many areas.
Our simulations show that we
expect larger yields than LHC-b in all charged particle final states,
at comparable or better signal-to-noise ratios, 
and we have large advantages in final states with photons.

For example, we compare the reaction $B^o\to\rho\pi$ in Table~\ref{rhopic0}.
 Both sets of numbers are calculated for
$10^7$ seconds at a luminosity of $2\times 10^{32}$ cm$^{-2}$s$^{-1}$.
We have corrected the LHC-b numbers by normalizing them to the
branching ratios used by BTeV.

\begin{table}[htp]
\begin{center}
\caption{Event yields and signal/background for $B^o\to\rho\pi$.}
\label{rhopic0}
\begin{tabular}{lccccc}\hline\hline
Mode & Branching Ratio &\multicolumn{2}{c}{BTeV} & \multicolumn{2}{c}{LHC-b}\\
     &             &   Yield  & S/B  & Yield & S/B \\\hline
$B^o\to\rho^{\pm}\pi^{\mp}$ & 2.8$\times 10^{-5}$ &9400 & 4.1 & 2140 & 0.8 \\
$B^o\to\rho^{o}\pi^{0}$ & 0.5$\times 10^{-5}$ &1350 & 0.3 & 880 & - \\
\hline\hline 
\end{tabular}
\end{center}
\end{table}

Furthermore, we intend to output on the order of 5 times more $b$'s per
second than \hbox{LHC-b} allowing for a greater range of physics studies. 
We also have the
capacity in our data acquisition system to accept a large number of 
directly produced charm decays.

\section{Conclusion}

BTeV is a powerful and precise scientific instrument capable of exquisite 
tests of the Standard Model. It has  great potential to discover new physics 
via rare or CP violating decays of heavy quarks.

\section*{Acknowledgements}
This work received partial support from  Fermilab which is operated by
University Research Association Inc. under Contract No. DE-AC02-76CH03000
with the United States Department of Energy.


\begin{thebibliography}{999}
\bibitem{proposal} The BTeV Proposal, which was submitted in May 2000,
may be found at \newline
           http://www-btev.fnal.gov/public\_documents/btev\_proposal/
\newline 
or may be accessed from the BTeV Homepage at
\newline
           http://www-btev.fnal.gov/btev.html
 
\bibitem{ckm} N. Cabibbo, \Journal{\PRL}{10}{531}{1963};
{ M. Kobayashi and K.~Maskawa}, {\it Prog. Theor. Phys.} {\bf 49}, 652 (1973).


\bibitem{KAL}
R. Aleksan, B. Kayser and D. London, Phys. Rev. Lett. 73 (1994) 18
(hep-ph/9403341).

\bibitem{bcx}
F. Abe \etal., ``Measurement of the $B$ Meson Differential Cross-Section in
$p\overline{p}$ collisions at $\sqrt{s}$ = 1.8 TeV," CDF/PUB/BOTTOM/PUBLIC/3759
submitted to ICHEP '96 and references therein; F. Abe \etal., 
\Journal{\PRL}{75}{1451}{1995}; 
R. Abbott \etal., ``The $b\overline{b}$ Production Cross Section and Angular
Correlations in $p\overline{p}$ collisions at $\sqrt{s}$ = 1.8 TeV," 
FERMILAB-Pub-99/144-E; S. Abachi \etal, \Journal{\PRL}{74}{3548}{1995}.

\bibitem{bcxD0}
D. Fein, ``Tevatron Results on $b$-Quark Cross Sections and Correlations,"
presented at Hadron Collider Physics (HCP99) (Bombay), January 1999. 





\end{thebibliography}
\end{document}